\documentclass[twocolumn,english,notitlepage,superscriptaddress,nofootinbib, twocolumn]{revtex4-1}
\usepackage[T1]{fontenc}
\usepackage[latin9]{inputenc}
\usepackage{bm}
\usepackage{bbold}
\usepackage{amsmath}
\usepackage{amssymb}
\usepackage{graphicx}
\usepackage{comment}
\usepackage{enumerate}
\usepackage{braket}

\makeatletter
\usepackage{babel}

\usepackage{hyperref}
\hypersetup{
    colorlinks=true,
    linkcolor=blue,
    citecolor=red,
    filecolor=blue,      
    urlcolor=blue
     }

\newcommand{\afffive}{Theoretical Sciences Visiting Program, Okinawa Institute of Science and Technology Graduate University, Onna-son, Okinawa 904-0495, Japan}

\makeatother

\usepackage{babel}
\begin{document}
\title{Geometry-controlled Onset of Inertial Drag in Granular Impact}
\author{Hollis Williams}
\affiliation{\afffive}

\begin{abstract}

The impact of solid intruders into granular media is commonly described by a combination of quasi-static resistance and an inertial drag force proportional to the square of the impact speed. While intruder geometry is known to influence force magnitudes, its role in controlling the onset of inertial drag has remained largely unexplored. Here we present systematic impact experiments using conical intruders spanning a wide range of apex angles. By measuring the peak acceleration during impact, we show that the emergence of a well-defined inertial response depends sensitively on cone geometry. Blunt cones exhibit quadratic scaling with impact speed over the full range of velocities studied, whereas sharper cones display a delayed transition to inertial behavior at higher speeds. We define a geometry-dependent crossover speed marking the onset of the inertial regime and find that it scales approximately linearly with the cone angle through $\tan\phi$. Once the inertial regime is established, the peak force collapses when rescaled by $\tan\phi$, indicating that cone geometry controls the effective momentum transfer to the grains. These results demonstrate that intruder geometry governs not only the magnitude of inertial drag, but also the impact speed at which it becomes dominant.

\end{abstract}
\maketitle

\section{Introduction}
 
When a solid object impacts a granular bed, the grains exhibit a collective behavior which differs from that of a conventional solid or fluid \cite{beh, meer}.  Instead, the intruder experiences a highly dissipative resistance force which depends on the speed of impact, the geometry and density of the projectile, and the properties of the grains \cite{royer, katsuragi, katsuragi2, royer2, goldman}.  Although the penetration depth and force scaling for spherical and cylindrical intruders is well understood, less is known about the instantaneous forces which act during the early stages of impact \cite{uehara,newhall, ambroso, brz, joubad, clark3}.  Of particular importance is the peak force, which determines the level of mechanical stress experienced by the projectile and governs its subsequent dynamics \cite{krizou, zheng, bcheng, athani}.  The physics of this peak force is complicated by the fact that granular materials transmit stress through intermittent, short-lived force chains \cite{peters, takahashi, hurley, kalwar}.  These networks form and collapse rapidly as grains rearrange, producing a fluctuating deceleration signal that can only be resolved with force measurements of a sufficiently high frequency \cite{clark, li}.  Optical tracking or differentiation of high-speed video images are widely used, but is typically unable to capture these rapid transients with precision \cite{best, best2}.

Several empirical models have been proposed to describe the resisting force during granular impact.  A widely used framework is the Katsuragi-Durian force law, which decomposes the stopping force into a depth-dependent frictional contribution and a velocity-dependent inertial term \cite{katsuragi, goldman}.  Whilst this model successfully captures the behavior of spherical and cylindrical intruders, its validity for non-axisymmetric or conical intruders remains unclear.  This is an important point, since simulations indicate that intruder geometry strongly influences how grains collide with the surface and how momentum is redirected into the granular bed, suggesting that the force response should depend sensitively on cone angle \cite{abram, yoon, ye}.
 
Cone-shaped intruders are also of interest because their pointed geometry concentrates collisions near the tip, with the effective projected area varying strongly with cone angle. As a result, a change in the angle of the tip modifies both the volume of grains displaced during the initial stages of impact and the distribution of force chains along the surface. A key open question is therefore whether there is a scaling for the peak force experienced by a conical projectile in terms of the cone angle.  Simulations have examined how frictional and inertial components of the drag force depend on the geometry of a conical intruder, but these studies primarily characterize the depth-dependent force terms during steady penetration \cite{zaidi}.

Experimental tests of this geometric dependence have been more limited and focus on the average drag force experienced by a cone-tipped cylinder, rather than the peak force \cite{ wang}.  Other studies of conical impacts have primarily inferred forces from trajectory measurements, which provide average dynamics but not detailed scaling of the peak deceleration with shape \cite{best}.  There is no previous systematic experimental study which investigates exactly how the peak impact force or the initial deceleration scales with cone angle.  In this work, we address this question by performing controlled impact experiments with a set of conical projectiles with varying half-angle and mass.  The projectile is fitted with a  high-bandwidth piezoelectric accelerometer, enabling us to resolve the force fluctuations during impact.  

By varying the impact speed, we obtain a data set that allows us to probe both the magnitude of the peak force and the conditions under which inertial drag becomes dominant. Our results show that, despite differences in cone geometry, the peak force in the inertial regime collapses onto a simple scaling form involving the impact velocity and a geometric factor depending on the cone half-angle. At the same time, cone geometry controls the impact speed at which this inertial response first emerges, revealing a geometry-dependent onset of the inertial regime. These findings indicate that the early-time force response is governed by collective inertial collisions with grains, whereas the cone angle sets the effective momentum flux into the material. Together, they provide a unified framework for comparing the dynamics of conical intruders impacting granular beds and connect naturally with existing collisional models of granular impact.



\section{Methods}


In this experiment, projectiles with an embedded 712-B3 EMIC accelerometer with range $\pm 510 g$ were released from rest above a granular bed.  To validate the accelerometer setup, we performed repeated drops of a steel sphere of diameter $D=2.6$ cm instrumented with an embedded accelerometer at heights $h$ between $8$ and 45 cm.  A hole was drilled in the top of the sphere and the accelerometer was fixed flat inside, with a notch drilled parallel so that the cable could exit horizontally.  A circular piece of thick tape was fixed above the accelerometer and the head of a light screw with base height 0.4 cm and 
thread length 0.6 cm was stuck to the tape  to enable drops at rest from an electromagnet.  The total combination of accelerometer, tape and screw added an additional 5 g to the mass of the sphere.  Although the cable is exposed in this setup, the peak force is expected to occur extremely quickly before the granular material can reach the cable at the top of the sphere and influence the acceleration signal.  It is also possible for the cable to fall on the sand before the sphere impacts, but this only provides a small amount of additional noise in the time trace prior to the peak force and does not affect the measured peak.



 The impact velocity was calculated as $v = \sqrt{2gh}$, with $\pm 0.5$ cm uncertainty in the drop height $h$ propagated to the horizontal error bars.  For each height, at least five drops were performed.  The bed was prepared each time by pouring into the container, rocking back and forth with decreasing amplitude, and lightly leveling with a straight edge.  Acceleration was recorded at 20 kHz over a time interval of 2 s. The primary deceleration peak was extracted from each trace, converted to units of $g$, and the mean across all drops was reported as the measured value, with vertical error bars corresponding to the standard deviation.  The measured accelerations scale approximately as $v^2$, as illustrated by the dashed red line in Fig. 1. This confirms that our setup and mounting procedure capture peak decelerations with sufficient reliability to obtain the expected scaling law.  It also confirmed that possible systematic offsets due to the mounting of the accelerometer did not affect the ability to extract the overall scaling.  This setup was applied in all subsequent experiments with cones.

The cones were printed using PETG at 100$\%$ infill (shown in Table 1).  The diameter of the cones was kept constant at $D =6$ cm and the cone half-angle $\phi$ was varied between 20 and 70$^{\circ}$.  To remove printing artifacts, all the cones were lightly sanded with fine 320-grit sandpaper and the tips were flattened to a nominal diameter between 1 and 2 mm.  This diameter is negligible relative to the cone diameter $D$ and is not expected to affect measured peak accelerations.  To ensure stability during free fall, the cones were taped flat to the bottom of a light wooden rod with length 13.2 cm and diameter 3.5 cm.  The rod was then dropped at rest from the electromagnet using a screw in the top, with the mass of the rod and accelerometer included in the total impactor mass.  A laser was also vertically aligned above the center of the impactor to monitor any tilting during impact.  For the blunt cones, the maximum drop height was restricted to avoid contact between the intruder assembly and the mounting notch. As a result, the highest impact speeds differ slightly between cone angles.

Glass beads with diameter between 300 and 425 $\mu$m were used for the granular material.  The depth of the granular bed was 16 cm which exceeds the maximum cone length in the study by at least a factor of two, hence measured peak accelerations occur well before possible boundary effects from the bottom of the container. 
The container was a polyethylene bucket with diameter 45 cm, height 17 cm, and wall thickness 0.3 cm.  The container diameter is 7.5 times larger than the projectile diameter. This ratio is comparable to or larger than those used in previous granular impact studies, where wall effects were found to be weak \cite{seguin, espinosa}. We therefore expect boundary effects from the walls to be negligible for the quantities of interest.

\begin{figure}

\includegraphics[width=80mm]{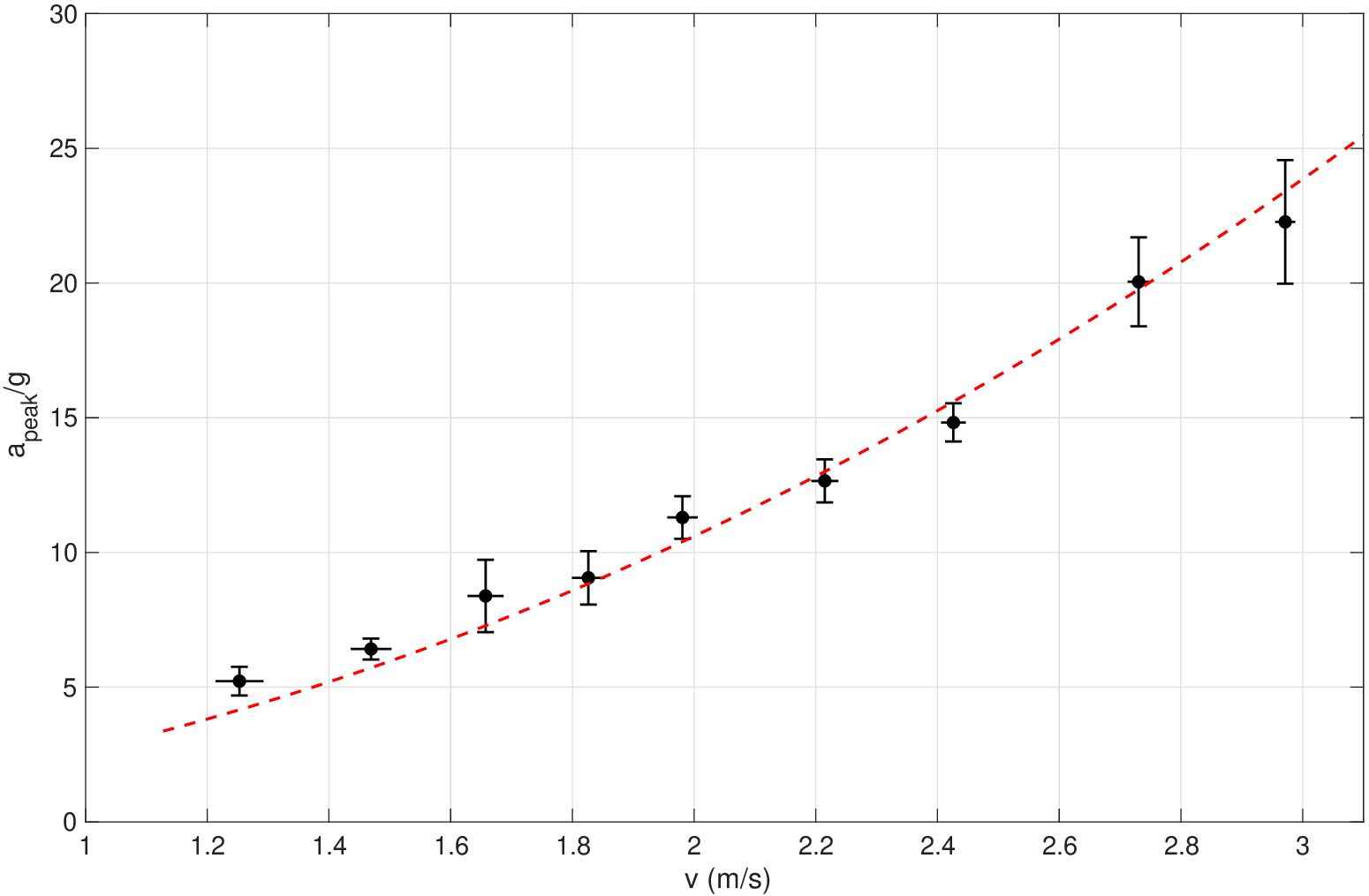}%

\caption{\label{fig:epsart} Peak acceleration $a_{\text{peak}}$ as a function of impact speed $v$ into glass beads for a steel sphere with diameter $D = 2.6$ cm.  Markers show mean values and error bars indicate standard deviation.  The dashed red line shows the known $a_{\text{peak}} \propto v^2$ scaling for sphere impacts (valid for $v\gtrsim$ 1.5 ms$^{-1}$) and is plotted as a reference to validate the accelerometer response.}
\end{figure}

\begin{table}[h!]
\centering
\caption{Impactors}
\begin{tabular}{lcc}
\hline
\textbf{Impactor} & \textbf{Diameter (cm)} & \textbf{Mass (g)} \\
\hline
Steel sphere & 2.6 & 81  \\
PETG cone, half-angle 70$^{\circ}$  & 6  & 102 \\
PETG cone, half-angle 60$^{\circ}$  & 6  & 108 \\
PETG cone, half-angle 50$^{\circ}$  & 6  & 118 \\
PETG cone, half-angle 40$^{\circ}$  & 6  & 129 \\
PETG cone, half-angle 30$^{\circ}$  & 6  & 147\\
PETG cone, half-angle 20$^{\circ}$  & 6  & 184 \\
\hline
\end{tabular}
\end{table}

\section{Emergence of Inertial Regime}

Granular impact dynamics is commonly described using a two-term force law in which the resisting force acting on an intruder is decomposed into a quasi-static contribution and an inertial drag proportional to the square of the intruder speed. In its standard form, the Katsuragi-Durian force law is written as

\begin{equation}
 F(z,v) = F_{\text{qs}} (z) + C \rho_g A v^2 ,   
\end{equation}

\noindent
where $F_{\text{qs}} (z)$ represents a depth-dependent resistance associated with frictional and gravitational loading of the grains, whilst the inertial term accounts for momentum transfer through grain-intruder collisions, with $A$ the effective area of the projectile and $\rho_{g}$ the grain density. This description has been shown to successfully capture the dynamics of spherical and blunt intruders once an inertial response is established.

An implicit assumption of this framework is that the impact response has entered a collective inertial regime in which momentum transfer to the grains dominates the dynamics.  Fig. 2 demonstrates that, for conical intruders, this assumption is strongly geometry dependent. Representative acceleration traces are shown for blunt (70$^{\circ}$) and sharp (30$^{\circ}$) cones impacting the granular bed at different impact speeds. For the blunt cone [left panel of Fig. 2], even relatively low impact speeds produce a smooth, reproducible acceleration peak characteristic of an inertial response. In contrast, at comparable speeds the sharp cone exhibits strongly intermittent acceleration signals dominated by discrete force chain events [inset of right panel of Fig. 2], and no inertial regime is apparent.

Only when the impact speed of the sharp cone is increased sufficiently does the response transition to a reproducible inertial form, with a well-defined acceleration peak similar to that observed for blunt cones [right panel of Fig. 2]. These observations indicate that cone angle delays the transition from quasi-static to inertial behavior.  For sharp cones, force transmission remains intermittent at low speeds, and the inertial regime emerges only beyond a cone angle-dependent impact speed. In this intermittent regime, the impact dynamics are dominated by localized force chain interactions rather than a collective inertial response.  Peak acceleration cannot be consistently determined for sharp cones in this pre-inertial regime, so these impacts are excluded from quantitative analysis.   

For the sharpest cones ($\phi = 20^{\circ}$ and $\phi = 30^{\circ}$), the force response at low and intermediate impact speeds is characterized by strongly fluctuating acceleration signals, reflecting the transmission of momentum through localized grain contacts rather than a spatially extended jamming front. As the impact speed increases, the inertial contribution to the resisting force grows and the response becomes progressively more collective, allowing a reproducible peak acceleration to be extracted for the $\phi = 30^{\circ}$ cone once inertial drag dominates.  Beyond this point, the inertial contribution to the resisting force grows rapidly with impact speed, leading to increasingly impulsive acceleration peaks that approach the dynamic range and mechanical limits of the accelerometer.  To avoid sensor saturation or mechanical interference, data collection for the 30° cone was therefore restricted to the minimum speed range required to establish inertial scaling and not extended further. For the $\phi = 20^{\circ}$ cone, impulsive force signals persist across the accessible speed range and no reproducible peak acceleration could be consistently extracted: this geometry was therefore excluded from further analysis.

\begin{figure}

\includegraphics[width=41mm]{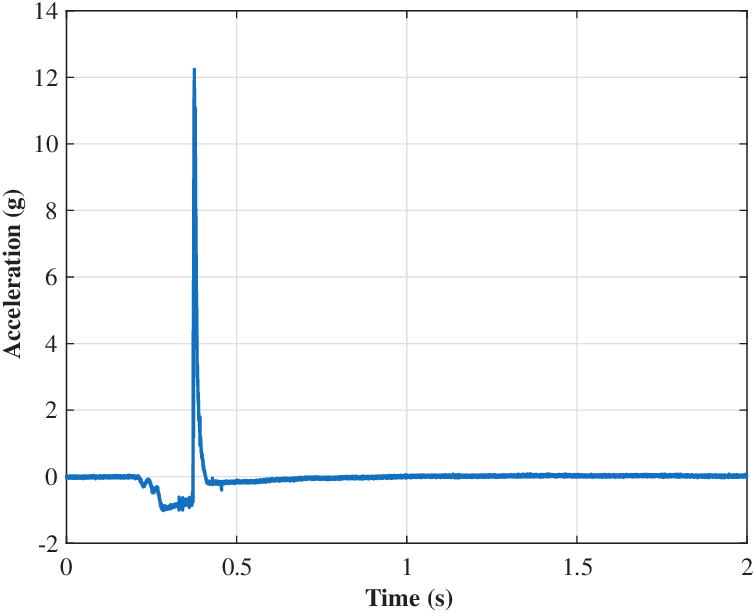}%
\includegraphics[width=39mm]{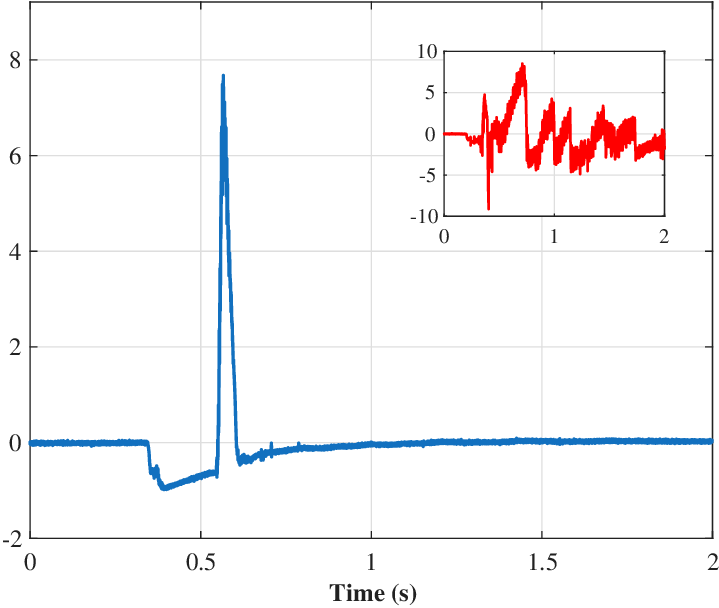}%

\caption{\label{fig:epsart}  Representative acceleration traces for conical intruders impacting a granular bed. Left panel: blunt $70^{\circ}$ cone with impact speed $v = 1.1$ ms$^{-1}$ showing a smooth, reproducible inertial peak. Right panel: sharp $30^{\circ}$ cone with impact speed $v = 1.8$ ms$^{-1}$ with an inertial peak.  Inset: same sharp cone with impact speed $v = 1.2$ ms$^{-1}$ exhibiting an intermittent acceleration signal associated with localized force chain events and failure to form a collective inertial response.  }
\end{figure}



The representative acceleration traces in Fig. 2 illustrate that the onset of a reproducible inertial response depends on cone geometry: blunt cones display clean inertial peaks at all speeds, whereas sharper cones exhibit intermittent, non-reproducible traces at low impact speeds. To quantify this transition, Fig. 3 shows the peak acceleration $a_{\text{peak}}$ as a function of $v^2$ for all cone angles. For blunt cones, $a_{\text{peak}}$ follows a quadratic dependence on $v$ over the full range of measured speeds. For sharper cones, deviations from quadratic scaling appear at low speeds, with a gradual crossover to inertial behavior at higher speeds. This observation motivates the definition of a crossover speed $v_c$, which marks the geometry-dependent onset of the inertial regime.

\begin{figure}

\includegraphics[width=80mm]{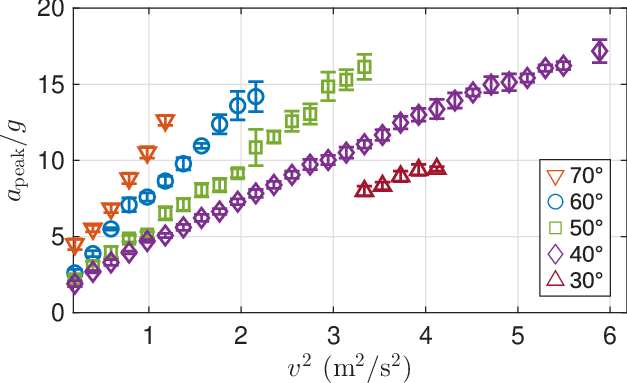}%

\caption{\label{fig:epsart}  Peak acceleration $a_{\text{peak}}$ as a function of impact speed squared $v^2$ for conical intruders with varying cone half-angle. Blunt cones exhibit quadratic scaling over the full speed range, while sharper cones deviate at low speeds and only approach quadratic scaling above a geometry-dependent crossover speed. This trend indicates that the onset of the inertial regime is delayed as cone angle is decreased. }
\end{figure}

To make this geometry-dependent crossover quantitative, we define a characteristic crossover speed $v_c$ as the lowest impact speed above which the peak acceleration remains consistent with the quadratic inertial scaling observed at higher speeds. Operationally, this scaling is identified from the high speed data for each cone, where $a_{\text{peak}} \propto v^2$ holds robustly.  The crossover speed $v_c$ is then estimated as the point below which the data systematically deviate from this inertial trend.

\begin{figure}

\includegraphics[width=80mm]{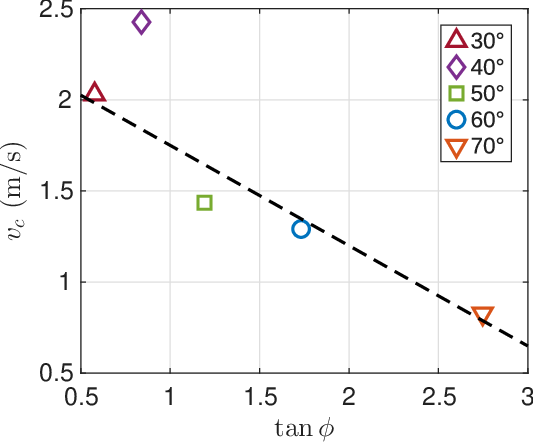}%

\caption{\label{fig:epsart} Geometry-dependent characteristic crossover speed $v_c$ associated with the emergence of inertial drag, plotted as a function of cone angle $\phi$.  For blunt and very sharp cones, the transition to inertial behavior is sharp and $v_c$ is well defined, varying approximately linearly with $\tan \phi$ (shown by the black dashed line). For intermediate cone angles, the crossover occurs gradually over a finite range of impact speeds, and the plotted value should be interpreted as an effective onset rather than a unique threshold.}
\end{figure}

Fig. 4 shows the resulting crossover speed $v_c$ plotted against $\tan \phi$.  For both very blunt and very sharp cones, $v_c$ is well defined and varies approximately linearly with 
$\tan \phi$, indicating that cone geometry controls the impact speed required for inertial drag to dominate. For intermediate cone angles, however, the crossover occurs gradually over a finite range of speeds rather than at a sharply defined threshold.  The non-monotonic behavior observed for intermediate cone angles reflects the absence of a sharply defined transition in this regime, rather than a breakdown of the underlying inertial scaling.  

We argue that the sharpness of the crossover is physically due to how grain-scale force transmission evolves with intruder geometry. For blunt cones, a large projected area engages many grains simultaneously, producing a collective inertial response even at low impact speeds and yielding a sharply defined $v_c$. In contrast, very sharp cones concentrate stress along narrow force chains, stabilizing a quasi-static, intermittent response which persists until a well-defined inertial threshold is exceeded. Once this threshold is crossed, inertial drag rapidly becomes dominant.  For intermediate cone angles, neither of theses limits applies cleanly: force chains coexist with partially mobilized grain motion over an extended range of speeds. As a result, the transition to inertial drag is gradual rather than abrupt, leading to a broader crossover region.

The dependence of $v_c$ on $\tan \phi$ suggests that the relevant geometric control parameter is not simply the projected area of the intruder, but the ratio of normal to tangential stress imposed on the granular bed. For a conical intruder, $\tan \phi$ characterizes the lateral deflection of grains relative to the penetration direction and thus the efficiency with which vertical momentum is redistributed into collective grain motion. Consistent with this interpretation, alternative geometric parameters such as $\sin \phi$ which primarily reflect trivial changes in projected area, do not produce a comparable collapse of the crossover speed in the two limits we have mentioned (see Fig. 5).

\begin{figure}

\includegraphics[width=80mm]{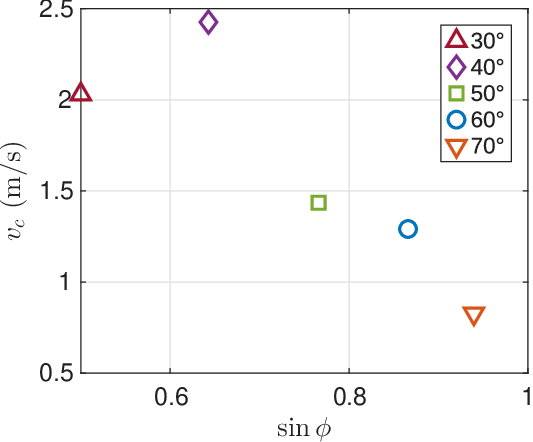}%

\caption{\label{fig:epsart} Geometry-dependent characteristic crossover speed $v_c$ associated with the emergence of inertial drag, plotted as a function of $\sin \phi$.  It can be seen that there is no collapse as with $\tan \phi$, hence the scaling of the onset to inertial drag cannot be explained by the change in projected area of the conical intruder.}
\end{figure}

\section{Peak force scaling}

Although the onset of inertial drag depends sensitively on cone geometry, once the inertial regime is reached the peak force is expected to obey a simpler and more universal scaling. We therefore examine the dependence of the peak impact force on impact speed and intruder geometry, focusing on data taken well beyond the crossover to inertial behavior.  The peak force experienced by the intruder follows directly from Newton's second law

\begin{equation}
  F_{\text{peak}} = m a_{\text{peak}},
\end{equation}

\noindent
where $m$ is the intruder mass and $a_{\text{peak}}$ is the maximum acceleration experienced during impact.  In granular impact, the inertial contribution to the resistive force is commonly written as

\begin{equation}
 F_{\text{inertial}} \sim C \rho A_{\text{eff}} v^2   
\end{equation}

\noindent
where $\rho$ is the bulk density of the granular medium, $A_{\text{eff}}$ is an effective area over which momentum is transferred, and $C$ is a dimensionless coefficient of order unity.  This form reflects momentum transfer to grains accelerated out of the intruder's path.

For a conical intruder, the effective area increases with penetration depth and is set by the cone geometry. To leading order, this geometric dependence enters via $\tan\phi$, which controls the lateral growth of the contact region and the volume of grains mobilized during impact. This motivates the rescaling

\begin{equation}
   \frac{F_{\text{peak}}}{\tan \phi} \sim v^2 
\end{equation}

\noindent
in the inertial regime.

\begin{figure}

\includegraphics[width=80mm]{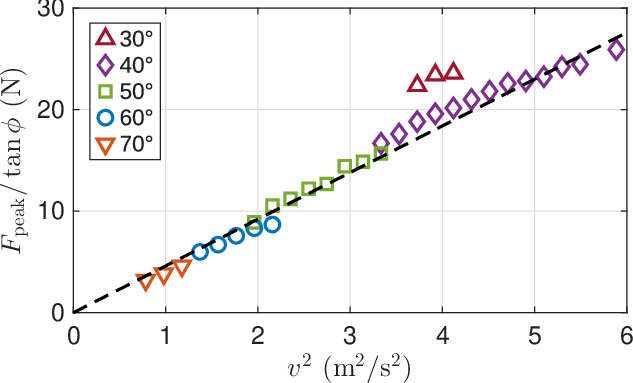}%

\caption{\label{fig:epsart} Peak force $F_{\text{peak}}$ normalized by $\tan\phi$ and plotted as a function of $v^2$ for all cone angles. Data for different intruder geometries collapse onto a single linear trend (shown with a dashed black line), consistent with inertial drag dominated by momentum transfer to the granular medium.  The slight upward deviation of the sharpest ($30^\circ$) cone likely reflects limited access to the fully developed inertial regime at the highest achievable impact speeds. }
\end{figure}


Fig. 6 shows $F_{\mathrm{peak}}/\tan\phi$ plotted as a function of $v^2$ for all cone angles. Only data points corresponding to the highest impact speeds for each cone are included. These points were chosen by only including speeds above a fixed fraction of the measured range, ensuring the measurements reflect fully developed inertial behavior.  Under this rescaling, the data collapse onto a single linear curve, demonstrating that once inertial drag dominates, the peak force is controlled by geometry in a simple and systematic way.  The sharpest ($30^\circ$) cone exhibits a small upward deviation from the common trend. This is likely due to the limited range of accessible impact speeds for this geometry, which restricts how deeply the system enters the fully developed inertial regime. 

Residual contributions from non-inertial force components may therefore still influence the measured peak force in this case. Importantly, this deviation does not undermine the overall collapse, but instead highlights the role of finite-speed effects near the onset of inertial behavior.  The excellent collapse of the peak force in the inertial regime demonstrates that once inertial drag is fully developed, the force response is reproducible and governed by a simple geometric prefactor. This contrasts with the geometry-dependent and often gradual onset of inertial behavior observed at lower impact speeds.

\section{Discussion}

A key lesson from the study of conical intruders on granular beds is that the force-bearing region ahead of an intruder is not determined solely by its projected area, but by the geometry-dependent structure of the force network amongst the grains and the associated jamming front. As a result, variations in cone angle probe changes in how momentum is transferred to the grains, as opposed to a trivial geometric rescaling.  Our measurements reveal two distinct and complementary roles of cone geometry. Firstly, the cone angle controls the impact speed at which inertial drag becomes experimentally observable. For blunt cones, a reproducible inertial response is present from the lowest speeds studied, whereas sharper cones exhibit intermittent, impulsive force transmission at low and intermediate speeds. In this regime, momentum transfer is mediated by transient force chains, and a well-defined global peak acceleration emerges only once the impact speed exceeds a geometry-dependent crossover. This delayed onset reflects the gradual inclusion of grains into a collective jamming front as the intruder penetrates more deeply and at higher speeds.

Secondly, once the inertial regime is established, the peak force exhibits robust scaling with respect to impact speed and cone angle.  The peak force collapses when rescaled by $\tan\phi$, indicating that cone geometry controls the effective momentum flux into the granular bed. This scaling is consistent with inertial drag arising from momentum transfer to a dynamically evolving, wedge-shaped region of grains whose lateral extent grows with cone angle.  The distinction between onset and magnitude is essential.  Although the inertial force law itself is universal, cone geometry governs when this regime becomes dominant and how efficiently momentum is coupled into the grains. For intermediate cone angles, we found that the transition between quasi-static and inertial response occurs gradually over a range of impact speeds, reflecting evolution of the force network rather than a sharp dynamical threshold.  These results demonstrate that intruder geometry plays a non-trivial role in granular impact, controlling both the emergence and the strength of inertial drag through its influence on jamming front formation and force chain organization. This geometry-dependent picture extends existing collisional models and provides a unified framework for comparing impacts across a wide range of intruder shapes.

\section*{Acknowledgments}

\noindent
This research was conducted whilst the author was visiting the Okinawa Institute of Science and Technology (OIST) through the Theoretical Sciences Visiting Program (TSVP).  The author thanks Tapan Sabuwala for useful discussions, Julio Manuel Barros for allowing use of his 3D printer, and Pinaki Chakraborty for allowing use of the laboratory space and equipment of the Fluid Mechanics Unit at OIST.






\end{document}